\documentclass[showpacs,preprintnumbers,amsmath,amssymb]{revtex4}

\usepackage{graphicx}
\usepackage{dcolumn}
\usepackage{bm}

\newcommand{\bq}{\begin{equation}}
\newcommand{\eq}{\end{equation}}
\newcommand{\bqa}{\begin{eqnarray}}
\newcommand{\eqa}{\end{eqnarray}}
\newcommand{\nn}{\nonumber \\}

\def\be     {\begin{equation}}
\def\ee     {\end{equation}}
\def\bea        {\begin{eqnarray}}
\def\eea        {\end{eqnarray}}
\def\bnn    {\begin{eqnarray*}}
\def\enn    {\end{eqnarray*}}

\begin{document}

\title{Critical field theory for ferromagnetic quantum criticality in the strong coupling regime of Hertz-Moriya-Millis theory}
\author{Ki-Seok Kim}
\affiliation{Department of Physics, POSTECH, Pohang, Gyeongbuk 790-784, Korea \\ Institute of Edge of Theoretical Science (IES), POSTECH, Pohang, Gyeongbuk 790-784, Korea}
\date{\today}

\begin{abstract} 
We develop U(1) slave spin-rotor theory, suggesting a metal-metal transition from Landau's Fermi-liquid state to a bad metal phase, as U(1) slave charge-rotor theory [Phys. Rev. B {\bf 70}, 035114 (2004)] describes a metal-insulator transition from Landau's Fermi-liquid state to a spin-liquid phase. U(1) slave spin-rotor formulation allows us to generalize Hertz-Moriya-Millis theory for ferromagnetic quantum phase transitions, replacing Landau's Fermi-liquid state with an incoherent metallic phase. As a result, we argue that localized magnetic moments emerge to govern quantum critical physics in bad metals.
\end{abstract}


\maketitle

\section{Introduction}

Hertz-Moriya-Millis theory is our standard critical field theory for quantum criticality from Fermi liquids, describing dynamics of local order parameter fluctuations coupled with renormalized electrons \cite{HMM}. Dynamics of Fermi surface fluctuations gives rise to Landau damping for order parameter fluctuations, resulting in the dynamical critical exponent $z$ larger than $1$, for example, $z = 2$ for antiferromagnetic quantum criticality and $z = 3$ for ferromagnetic quantum criticality \cite{QCP_Review1,QCP_Review2}. As a result, the critical field theory turns out to live above the upper critical dimension, identifying a quantum critical point with a Gaussian fixed point. However, there exist dangerously irrelevant interaction vertices, responsible for breaking the hyperscaling relation, which does not allow $\omega/T$ scaling \cite{QCP_Review1,QCP_Review2}, where $\omega$ is frequency and $T$ is temperature. In particular, ferromagnetic quantum phase transitions from Landau's Fermi-liquids have been proposed to be generically of first order instead of second order, where correlation effects between ferromagnetic spin fluctuations, given by Fermi-surface fluctuations and described by some types of vertex corrections, turn out to cause non-analytic contributions for momentum with a negative coefficient in the uniform spin susceptibility, which invalidates the Hertz-Moriya-Millis description for ferromagnetic quantum criticality \cite{BKV,Pepin_FM,Chubukov_FM}.

In this study we revisit the Hertz-Moriya-Millis theory for ferromagnetic quantum phase transitions. While the Hertz-Moriya-Millis theory deals with effective interactions between renormalized electrons and ferromagnetic spin fluctuations perturbatively, we take them into account non-perturbatively, ``diagonalizing" the spin-fermion coupling term first \cite{FMLQCP,Kim_U1SSR_AFQCP,With_MDKim1,With_MDKim2}. This strong coupling approach is to take the ``spherical" coordinate instead of the ``cartesian" coordinate in the order parameter space, where ferromagnetic spin fluctuations are decomposed into longitudinal (amplitude) and transverse (directional) spin fluctuations. Actually, an effective theory based on the spherical coordinate of the order parameter space has been proposed for metal-insulator transitions, referred to as U(1) slave charge-rotor theory \cite{U1SCR}, describing charge dynamics in terms of the density-phase representation. Here, we apply the scheme of the U(1) slave charge-rotor theory to spin dynamics, referred to as U(1) slave spin-rotor theory, where the Hertz-Moriya-Millis theory for critical ferromagnetic spin fluctuations is reformulated in terms of critical longitudinal and transverse spin excitations \cite{FMLQCP,Kim_U1SSR_AFQCP}. The U(1) slave spin-rotor representation reformulates the Hertz-Moriya-Millis theory in terms of renormalized electrons, longitudinal and transverse spin fluctuations, gapless spin singlet excitations (gauge fluctuations), and their interactions. Introducing quantum corrections into this effective field theory based on the Eliashberg approximation, we find a renormalized field theory, where both longitudinal spin fluctuations and gauge fluctuations are described by $z = 3$ critical dynamics due to Landau damping as expected but dynamics of transverse spin fluctuations is characterized by nonlocal interactions in time, which originate from renormalization by the $z = 3$ critical spin dynamics. Performing the scaling analysis for the resulting renormalized field theory, we find a critical field theory in the strong coupling regime of the Hertz-Moriya-Millis theory, which looks quite different from the Hertz-Moriya-Millis theory at least in this level of approximation. This implies the existence of a novel fixed point in the strong coupling regime of the Hertz-Moriya-Millis theory. An essential feature of this fixed point is that nonlocal interactions in dynamics of transverse spin fluctuations give rise to localization for the dynamics of transverse spin fluctuations, which turns out to be responsible for the fact that more interaction vertices are marginal, involved with transverse spin fluctuations, while only the spin-fermion interaction vertex is marginal in the Hertz-Moriya-Millis theory. It is quite appealing that the critical field theory differs from the Hertz-Moriya-Millis theory even in three dimensions for the case of the strong coupling regime, where locally critical transverse spin fluctuations modify the Hertz-Moriya-Millis theory.

The underlying physical picture of the U(1) slave spin-rotor theory is as follows. As the U(1) slave charge-rotor theory describes one metal-insulator Mott transition from Landau's Fermi-liquid state to a spin-liquid phase \cite{U1SCR}, the U(1) slave spin-rotor theory suggests one metal-metal transition from Landau's Fermi-liquid state to a bad metal phase, where the coherence of the electron quasiparticle nature disappears through scattering between renormalized electrons and emergent localized magnetic excitations \cite{U1SSR}. Suppose a Hubbard-type minimal model. It is natural to expect an insulating phase when the strength of electron correlations exceeds a critical value. Our interesting region is a metallic state near the metal-insulator transition. Physically, one may speculate that dynamics of renormalized electrons becomes localized near the metal-insulator transition, which gives rise to enhancement of density of states near the Fermi energy. Then, we expect that ferromagnetic correlations can be enhanced, considering the Stoner criteria \cite{Many_Body_Textbook}. Of course, it is not clear at all whether we can apply the Stoner picture to this regime. In particular, such ferromagnetic correlations are expected to compete with enhanced antiferromagnetic fluctuations, which result from stronger electron correlations. In other words, frustration may appear in the metallic regime near the Mott transition as a result of the competition between ferromagnetic and antiferromagnetic interactions. Our physical picture is that almost localized magnetic moments arise around this regime, responsible for strong inelastic scattering between electron quasiparticle excitations. As a result, an incoherent bad metallic state is proposed to arise near the Mott transition. This scenario reminds us of the dynamical mean-field theory description \cite{DMFT_Bad_Metal}, where localized magnetic moments are introduced explicitly by hands. The present ferromagnetic quantum criticality is suggested to appear in the bad metallic phase, which should be distinguished from the conventional Stoner instability in the Landau's Fermi-liquid state.

\section{Review on U(1) slave spin-rotor theory}

\subsection{CP$^{1}$ representation}

We start from a Hubbard-type model \begin{eqnarray} && Z = \int D c_{i\sigma} \exp\Bigl[ - \int_{0}^{\beta} d \tau \Bigl\{ \sum_{i} c_{i\sigma}^{\dagger} (\partial_{\tau} - \mu) c_{i\sigma} - t \sum_{ij} (c_{i\sigma}^{\dagger} c_{j\sigma} + H.c.) + \frac{g}{2} \sum_{i} c_{i\uparrow}^{\dagger} c_{i\uparrow} c_{i\downarrow}^{\dagger} c_{i\downarrow} \Bigr\} \Bigr] , \end{eqnarray} where $c_{i\sigma}$ is an electron field with spin $\sigma$ at site $i$ and $\mu$, $t$, and $g$ are its chemical potential, hopping parameter, and interaction strength, respectively. Summation for the spin index is omitted for a simple notation.

Performing the Hubbard-Stratonovich transformation involved with ferromagnetic instability, we obtain \begin{eqnarray} && Z = \int D c_{i\sigma} D \boldsymbol{\Phi}_{i} \exp\Bigl[ - \int_{0}^{\beta} d \tau \Bigl\{ \sum_{i} c_{i\sigma}^{\dagger} (\partial_{\tau} - \mu) c_{i\sigma} - t \sum_{ij} (c_{i\sigma}^{\dagger} c_{j\sigma} + H.c.) - \sum_{i} c_{i\alpha}^{\dagger} \boldsymbol{\Phi}_{i} \cdot \boldsymbol{\sigma}_{\alpha\beta} c_{i\beta} + \frac{1}{2g} \sum_{i} \boldsymbol{\Phi}_{i}^{2} \Bigr\} \Bigr] , \label{FM_HS} \end{eqnarray} where only the particle-hole sector $\boldsymbol{\Phi}_{i}$ in the spin-triplet channel, corresponding to magnetization, is taken into account and other interactions are assumed to be not critical.

Resorting to the CP$^{1}$ representation \cite{Spin_Textbook}, the magnetization order parameter is expressed as follows \begin{eqnarray} && \boldsymbol{\Phi}_{i} \cdot \boldsymbol{\sigma}_{\alpha\beta} = \phi_{i} U_{i\alpha\gamma} \sigma^{3}_{\gamma\delta} U_{i\delta\beta}^{\dagger} , \label{CP1} \end{eqnarray} where the scalar component of $\phi_{i}$ is an amplitude-fluctuation field and the SU(2) matrix of $\boldsymbol{U}_{i} = \left(\begin{array}{cc} z_{i\uparrow} & z_{i\downarrow}^{\dagger} \\ z_{i\downarrow} & - z_{i\uparrow}^{\dagger} \end{array} \right)$ is a directional-fluctuation field.

Inserting Eq. (\ref{CP1}) into Eq. (\ref{FM_HS}) and introducing the projective representation \begin{eqnarray} && c_{i\alpha} = U_{i\alpha\beta} f_{i\beta} , \end{eqnarray} where $z_{i\sigma}$ is a bosonic spinon field to describe directional fluctuations of spins and $f_{i\sigma}$ is a fermionic holon field to describe dynamics of renormalized electrons, we obtain \begin{eqnarray} && Z = \int D f_{i\alpha} D U_{i\alpha\beta} D \phi_{i} \exp\Bigl[ - \int_{0}^{\beta} d \tau \Bigl\{ \sum_{i} f_{i\alpha}^{\dagger} [(\partial_{\tau} - \mu) \delta_{\alpha\beta} - U_{i\alpha\gamma}^{\dagger} \partial_{\tau} U_{i\gamma\beta} ] f_{i\beta} \nn && - t \sum_{ij} (f_{i\alpha}^{\dagger} U_{i\alpha\gamma}^{\dagger} U_{j\gamma\beta} f_{j\beta} + H.c.) - \sum_{i} \phi_{i} f_{i\alpha}^{\dagger} \sigma^{3}_{\gamma\delta} f_{i\beta} + \frac{1}{2g} \sum_{i} \phi_{i}^{2} \Bigr\} \Bigr] , \label{CP1_action} \end{eqnarray} where no approximations have been made.

Benchmarking the U(1) slave charge-rotor theory \cite{U1SCR} and following the procedure of refs. \cite{FMLQCP,Kim_U1SSR_AFQCP,With_MDKim1,With_MDKim2}, we reach the following expression as our starting point \begin{eqnarray} && Z_{SR} = \int D f_{i\sigma} D z_{i\sigma} D \phi_{i} D \lambda_{i} e^{- \int_{0}^{\beta} d \tau L_{eff}} , ~~~~~ L_{eff} = L_{f} + L_{z} + L_{c} , \nn && L_{f} = \sum_{i} f_{i\sigma}^{\dagger} (\partial_{\tau} - \mu - \sigma \phi_{i}) f_{i\sigma} - t \sum_{ij} ( f_{i\sigma}^{\dagger} \chi_{ij}^{f} f_{j\sigma} + H.c ) , \nn && L_{z} = \frac{1}{2g} \sum_{i} \Bigl( z_{i\sigma}^{\dagger} \partial_{\tau} z_{i\sigma} - \frac{\phi_{i}}{2} \Bigr)^{2} - t \sum_{ij} ( z_{i\sigma}^{\dagger} \chi_{ij}^{z} z_{j\sigma} + H.c. ) + i \sum_{i} \lambda_{i} (|z_{i\sigma}|^{2} - 1) , \nn && L_{c} = t \sum_{ij} ( \chi_{ij}^{f} \chi_{ij}^{z} + H.c. ) , \label{U1SSR} \end{eqnarray} referred to as U(1) slave spin-rotor (SR) theory. $\chi_{ij}^{f}$ and $\chi_{ij}^{z}$ are hopping parameters for holons (renormalized electrons) and spinons (directional fluctuations), respectively, which arise from the Hubbard-Stratonovich transformation of the hopping term of $- t \sum_{ij} (f_{i\alpha}^{\dagger} U_{i\alpha\gamma}^{\dagger} U_{j\gamma\beta} f_{j\beta} + H.c.)$ in Eq. (\ref{CP1_action}). $\lambda_{i}$ is a Lagrange multiplier field to impose the slave-rotor constraint.

In order to compare the U(1) slave spin-rotor theory with the U(1) slave charge-rotor theory, we would like to recall the U(1) slave charge-rotor representation of the Hubbard model \cite{U1SCR} \begin{eqnarray} && Z_{CR} = \int D f_{i\sigma} D b_{i} D \varphi_{i} D \lambda_{i} e^{- \int_{0}^{\beta} d \tau L_{eff}} , ~~~~~ L_{eff} = L_{f} + L_{b} + L_{c} , \nn && L_{f} = \sum_{i} f_{i\sigma}^{\dagger} (\partial_{\tau} - \mu - i \varphi_{i}) f_{i\sigma} - t \sum_{ij} ( f_{i\sigma}^{\dagger} \chi_{ij}^{f} f_{j\sigma} + H.c ) , \nn && L_{b} = - \frac{1}{2g} \sum_{i} ( b_{i}^{\dagger} \partial_{\tau} b_{i} - i \varphi_{i} )^{2} - t \sum_{ij} ( b_{i}^{\dagger} \chi_{ij}^{b} b_{j} + H.c. ) + i \sum_{i} \lambda_{i} (|b_{i}|^{2} - 1) , \nn && L_{c} = t \sum_{ij} ( \chi_{ij}^{f} \chi_{ij}^{b} + H.c. ) , \label{U1SCR} \end{eqnarray} referred to as the U(1) slave charge-rotor (CR) theory, where an electron field is decomposed as $c_{i\sigma} = b_{i}^{\dagger} f_{i\sigma}$ with the rotor constraint of $|b_{i}|^{2} = 1$.

As can be seen, the U(1) slave spin-rotor theory looks quite similar to the U(1) slave charge-rotor theory. However, the U(1) slave spin-rotor theory is not consistent in contrast with the U(1) slave charge-rotor theory. The positive sign in $\frac{1}{2g} \sum_{i} \Bigl( z_{i\sigma}^{\dagger} \partial_{\tau} z_{i\sigma} - \frac{\phi_{i}}{2} \Bigr)^{2}$ favors stronger directional fluctuations while it is negative in the U(1) slave charge-rotor theory, serving a parabolic potential for charge fluctuations and guaranteeing the stability of their dynamics. This difference originates from the opposite sign when the Hubbard-$g$ term is decomposed into charge and spin channels.

One may ask why the slave spin-rotor theory is given by U(1) gauge theory instead of SU(2). Generally speaking, an effective field theory in the spin-rotor representation is formulated as an SU(2) gauge theory. In the present study we keep only the third component of the gauge field with $\tau^{3}$, where $\tau^{3}$ is the third component of the pauli matrix. Then, the effective magnetic field as an order parameter field in the Hubbard-Stratonovich transformation is identified with the time component of the U(1) gauge field in the U(1) slave spin-rotor theory as the electric potential field is given by the time component of the U(1) gauge field in the U(1) slave charge-rotor theory. Fluctuations of off-diagonal components of SU(2) gauge fields are neglected, where they play their roles of emergent spin-orbit interactions, which flip ``spins" (spin quantum numbers) during the propagation of elementary excitations. We would like to recall the SU(2) slave charge-rotor theory \cite{SU2SCR}, regarded to be an SU(2) generalization of the U(1) slave charge-rotor theory, where fluctuations of off-diagonal components of SU(2) gauge fields correspond to pairing fluctuations (singlet channel) associated with superconductivity. As the U(1) slave charge-rotor theory considers some limited regions of the SU(2) slave charge-rotor theory, we have considered the U(1) slave spin-rotor theory as the first step of the SU(2) slave spin-rotor theory. We speculate that this formulation may lead us to interacting topological states since the SU(2) slave spin-rotor formulation introduces effective spin-orbit interactions naturally.

\subsection{Role of amplitude fluctuations}

In order to cure the inconsistency of the U(1) slave spin-rotor theory, an idea is to introduce quantum corrections into the spinon dynamics in the random phase approximation (RPA), given by \begin{eqnarray} && S_{z} = \int_{0}^{\beta} d \tau \Bigl\{ \frac{1}{2g} \sum_{i} ( z_{i\sigma}^{\dagger} \partial_{\tau} z_{i\sigma} )^{2} - \frac{1}{8 g^{2}} \int_{0}^{\beta} d \tau' \sum_{i} \sum_{j} ( z_{i\sigma}^{\dagger} \partial_{\tau} z_{i\sigma} )_{\tau} \Bigl( \frac{1}{4g} \boldsymbol{I} - \boldsymbol{\Pi} \Bigr)^{-1}_{\tau\tau',ij} ( z_{j\sigma'}^{\dagger} \partial_{\tau'} z_{j\sigma'} )_{\tau'} \nn && - t \sum_{ij} ( z_{i\sigma}^{\dagger} \chi_{ij}^{z} z_{j\sigma} + H.c. ) + i \sum_{i} \lambda_{i} (|z_{i\sigma}|^{2} - 1) \Bigr\} . \label{RPA_Boson_Sector} \end{eqnarray} Here, \begin{eqnarray} && \Pi(\boldsymbol{q},i\Omega;m) = \frac{N_{\sigma}}{\beta} \sum_{i\omega} \sum_{\boldsymbol{k}} g_{f}(\boldsymbol{k}+\boldsymbol{q},i\omega+i\Omega) g_{f}(\boldsymbol{k},i\omega) \nonumber \end{eqnarray} is a polarization function given by a fermion-bubble diagram, regarded as the self-energy for amplitude fluctuations, where $g_{f}(\boldsymbol{k},i\omega)$ is the holon Green's function. $N_{\sigma}$ represents the spin degeneracy, which extends the number of spin degrees of freedom from $\sigma = \uparrow, \downarrow$ to $\sigma = 1, ..., N_{\sigma}$ in the $Sp(N_{\sigma}/2)$ representation \cite{Sp_N}.

This RPA-corrected spinon action is simplified further as follows \begin{eqnarray} && S_{z} = \int_{0}^{\beta} d \tau \Bigl[ - \int_{0}^{\beta} d \tau' \sum_{i} \sum_{j} ( z_{i\sigma}^{\dagger} \partial_{\tau} z_{i\sigma} ) \Bigl\{ \frac{\Pi(\boldsymbol{q},i\Omega)}{1 - 4 g \Pi(\boldsymbol{q},i\Omega)} \Bigr\}_{\tau\tau',ij} ( z_{j\sigma'}^{\dagger} \partial_{\tau'} z_{j\sigma'} ) \nn && - t \sum_{ij} ( z_{i\sigma}^{\dagger} \chi_{ij}^{z} z_{j\sigma} + H.c. ) + i \sum_{i} \lambda_{i} (|z_{i\sigma}|^{2} - 1) \Bigr] . \label{U1SSR_Spinon_Sector} \end{eqnarray} An essential modification is that the positive sign in the time-fluctuation part of Eq. (\ref{U1SSR}) turns into the negative sign as long as $1 - 4 g \Pi(\boldsymbol{q} \rightarrow 0,i\Omega \rightarrow 0) \geq 0$ which corresponds to a paramagnetic state. As a result, the U(1) slave spin-rotor theory becomes consistent for the description of spin dynamics.

\section{Renormalization group analysis}

\subsection{An effective field theory in the U(1) slave spin-rotor representation}

\begin{figure}[t]
\includegraphics[width=0.8\textwidth]{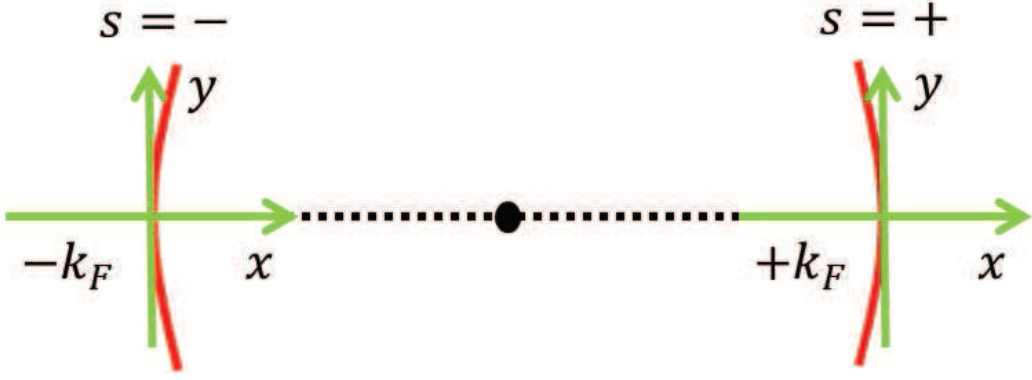}
\caption{A schematic diagram of a Fermi surface in the double patch construction. Red-curved lines denote a pair of Fermi surfaces, connected by $2\bm{k}_{F}$, inside which electrons are filled with. A coordinate system is defined as the figure for each patch of $s = \pm$.} \label{Double_Patch_Construction}
\end{figure}

Constructing an effective field theory in the U(1) slave spin-rotor representation and performing renormalization group analysis, we investigate the nature of ferromagnetic quantum criticality in the strong coupling regime of the Hertz-Moriya-Millis theory. Following the patch construction of refs. \cite{SSL_U1GT,Metlitski_U1GT} shown in Fig. \ref{Double_Patch_Construction}, we write down an effective field theory of the U(1) slave spin-rotor representation \bqa && Z = \int D f_{s\sigma} D z_{\sigma} D \phi D a \exp\Bigl[ - \int_{0}^{\beta} d \tau \int_{-\infty}^{\infty} d x \int_{-\infty}^{\infty} d y \Bigl\{ f_{s\sigma}^{\dagger} \Bigl(\partial_{\tau} - i s v_{F} \partial_{x} - \frac{v_{F}}{2\gamma} \partial_{y}^{2}\Bigr) f_{s\sigma} \nn && + \phi \Bigl( \partial_{\tau} - v_{\phi}^{2} \partial_{x}^{2} - v_{\phi}^{2} \partial_{y}^{2} + m_{\phi}^{2} \Bigr) \phi + \frac{u_{\phi}}{2} \phi^{4} + a ( - \partial_{\tau}^{2} - v_{a}^{2} \partial_{x}^{2} - v_{a}^{2} \partial_{y}^{2} ) a + z_{\sigma}^{\dagger} ( - v_{z}^{2} \partial_{x}^{2} - v_{z}^{2} \partial_{y}^{2} + m_{z}^{2} ) z_{\sigma} + \frac{u_{z}}{2} |z_{\sigma}|^{4} \nn && - g_{\phi} \phi \sigma f_{s\sigma}^{\dagger} f_{s\sigma} - e_{f} s v_{F} a \sigma f_{s\sigma}^{\dagger} f_{s\sigma} - g_{z} \phi z_{\sigma}^{\dagger} \partial_{\tau} z_{\sigma} - i e_{z} a [ z_{\sigma}^{\dagger} (\partial_{x} z_{\sigma}) - (\partial_{x} z_{\sigma}^{\dagger}) z_{\sigma}] \nn && - \int_{0}^{\beta} d \tau' \int_{-\infty}^{\infty} d x' \int_{-\infty}^{\infty} d y' ( z_{\sigma}^{\dagger} \partial_{\tau} z_{\sigma} )_{\bm{r},\tau} \Bigl( \frac{\Pi(\boldsymbol{q},i\Omega)}{1 - 4 g_{\phi} \Pi(\boldsymbol{q},i\Omega)} \Bigr)_{\bm{r}\bm{r}',\tau\tau'} ( z_{\sigma'}^{\dagger} \partial_{\tau'} z_{\sigma'} )_{\bm{r}',\tau'} \Bigr\} \Bigr] , \eqa regarded to be a continuum version of Eq. (\ref{U1SSR}), where the spinon sector is replaced with Eq. (\ref{U1SSR_Spinon_Sector}). $f_{s\sigma}$ is a low-energy renormalized electron field (holon) with spin $\sigma$ on the Fermi surface of a $s = \pm$ patch. Its dispersion relation is given by $\epsilon(k_{\parallel},k_{\perp}) = s v_{F} k_{\parallel} + \frac{v_{F}}{2\gamma} k_{\perp}^{2}$, where $k_{\parallel}$ is the longitudinal momentum out of the Fermi surface and $k_{\perp}$ is the transverse momentum along the Fermi surface. See Fig. 1. $v_{F}$ is a Fermi velocity and $\gamma$ is a Landau-damping coefficient \cite{U1GT_Scaling}. $\phi$ represents an amplitude-fluctuation field in the ferromagnetic channel whose dispersion relation is given by the non-relativistic spectrum of  $E_{\phi}(k_{\parallel},k_{\perp}) = v_{\phi}^{2} (k_{\parallel}^{2} + k_{\perp}^{2}) + m_{\phi}^{2}$, where this bare dispersion is not much relevant for its renormalized dynamics. $u_{\phi}$ denotes a mode-mode coupling constant. $a$ is a transverse gauge field with the relativistic dispersion $E_{a}(k_{\parallel},k_{\perp}) = v_{a} \sqrt{ k_{\parallel}^{2} + k_{\perp}^{2} }$, describing phase (transverse) fluctuations ($a_{ij}$) of the hopping parameter given by $\chi_{ij}^{f} = \chi^{f} e^{i \sigma a_{ij}}$ and $\chi_{ij}^{z} = \chi^{z} e^{i a_{ij}}$, where amplitude (longitudinal) fluctuations are assumed to be gapped, not relevant. $z_{\sigma}$ is a transverse spin-fluctuation field (spinon), where the temporal part is given by the one-loop correction from critical ferromagnetic amplitude fluctuations. $v_{z}$ and $m_{z}$ are the velocity and mass of spinons, respectively. We show that both the velocity and mass of spinons become renormalized to vanish at the ferromagnetic quantum critical point of $m_{\phi} = 0$, which may be identified with local quantum criticality. $u_{z}$ is the self-interaction parameter of spinons. Critical ferromagnetic amplitude fluctuations couple to both holons and spinons with coupling constants of $g_{\phi}$ and $g_{z}$, respectively, and gapless gauge fluctuations do to both holons and spinons with $e_{f}$ and $e_{z}$, respectively.

Next, we introduce quantum corrections into this field theory within the Eliashberg approximation, where $\sigma = \uparrow, \downarrow$ is generalized to $\sigma = 1, 2, ..., N_{\sigma}$ as discussed before. Both critical ferromagnetic amplitude fluctuations and gapless gauge fluctuations give rise to the $|\omega|^{\frac{2}{3}}$ self-energy correction with numerical constants in holon dynamics, originating from the $z = 3$ critical dynamics, where $z$ is the dynamical critical exponent \cite{QCP_Review1,QCP_Review2}. The polarization bubble of $\Pi(\boldsymbol{q},i\Omega)$ gives rise to Landau damping for both ferromagnetic amplitude and gauge fluctuations, where $\gamma$ is a damping coefficient. Critical ferromagnetic amplitude fluctuations give rise to not only the consistency for the dynamics of transverse spin fluctuations as discussed in the last section but also nonlocal correlations in their temporal dynamics responsible for anomalous scaling in various interaction vertices involved with transverse spin fluctuations. They also generate the holon-spinon coupling term, which turns out to play an important role in our ferromagnetic quantum phase transition, modifying the Hertz-Moriya-Millis theory. 

Performing the Fourier transformation toward the real space, we find a consistent U(1) slave spin-rotor effective field theory in terms of renormalized electrons, critical longitudinal spin (ferromagnetic amplitude) fluctuations, transverse spin fluctuations, U(1) gauge fluctuations, and their interactions, where quantum corrections are taken into account in the Eliashberg approximation near ferromagnetic quantum criticality \bqa && Z = \int D f_{s\sigma} D z_{\sigma} D \phi D a \exp\Bigl[ - \int_{0}^{\beta} d \tau \int_{-\infty}^{\infty} d x \int_{-\infty}^{\infty} d y \Bigl\{ f_{s\sigma}^{\dagger} \Bigl(- i \frac{c}{N_{\sigma}} (- \partial_{\tau}^{2})^{\frac{1}{3}} - i s v_{F} \partial_{x} - \frac{v_{F}}{2\gamma} \partial_{y}^{2} \Bigr) f_{s\sigma} \nn && + \phi \Bigl( \gamma \frac{\sqrt{- \partial_{\tau}^{2}}}{\sqrt{- \partial_{y}^{2}}} - v_{\phi}^{2} \partial_{y}^{2} \Bigr) \phi + \frac{u_{\phi}}{2} \phi^{4} + a \Bigl( \gamma \frac{\sqrt{- \partial_{\tau}^{2}}}{\sqrt{- \partial_{y}^{2}}} - v_{a}^{2} \partial_{y}^{2} \Bigr) a- \frac{g_{\phi}}{\sqrt{N_{\sigma}}} \phi \sigma f_{s\sigma}^{\dagger} f_{s\sigma} - \frac{e_{f}}{\sqrt{N_{\sigma}}} s v_{F} a \sigma f_{s\sigma}^{\dagger} f_{s\sigma} \nn && - \frac{g_{c}}{N_{\sigma}} \sigma f_{s\sigma}^{\dagger} f_{s\sigma} \frac{1}{\gamma \frac{\sqrt{- \partial_{\tau}^{2}}}{\sqrt{- \partial_{y}^{2}}} - v_{\phi}^{2} \partial_{y}^{2}} (z_{\sigma'}^{\dagger} \partial_{\tau} z_{\sigma'}) - \frac{g_{d}}{N_{\sigma}} ( z_{\sigma}^{\dagger} \partial_{\tau} z_{\sigma} ) \frac{1}{\gamma \frac{\sqrt{- \partial_{\tau}^{2}}}{\sqrt{- \partial_{y}^{2}}} - v_{\phi}^{2} \partial_{y}^{2}} ( z_{\sigma'}^{\dagger} \partial_{\tau} z_{\sigma'} ) \nn && + z_{\sigma}^{\dagger} ( - v_{z}^{2} \partial_{y}^{2} + m_{z}^{2} ) z_{\sigma} + \frac{u_{z}}{2} |z_{\sigma}|^{4} - \frac{g_{z}}{\sqrt{N_{\sigma}}} \phi z_{\sigma}^{\dagger} \partial_{\tau} z_{\sigma} - i \frac{e_{z}}{\sqrt{N_{\sigma}}} a [ z_{\sigma}^{\dagger} (\partial_{x} z_{\sigma}) - (\partial_{x} z_{\sigma}^{\dagger}) z_{\sigma}] \Bigr\} \Bigr] . \label{U1SSR_EFT} \eqa We would like to emphasize that all non-analytic operator expressions can be well defined in the momentum and frequency space, where this real-space expression should be regarded to be formal, introduced to give some insight with simplicity in the presentation. $c$ is a positive numerical constant, and $g_{c} \sim g_{\phi} g_{z}$ and $g_{d} \sim g_{z}^{2}$ are positive coupling constants. Resorting to robustness of the Fermi surface, we keep dynamics along the transverse momentum for boson excitations \cite{SSL_U1GT}. In other words, boson dynamics along $- \partial_{x}^{2}$ are not relevant.

We would like to point out an interesting aspect of the U(1) slave spin-rotor effective field theory [Eq. (\ref{U1SSR_EFT})]. As discussed in the introduction, correlation effects between ferromagnetic spin fluctuations invalidate the Hertz-Moriya-Millis description for ferromagnetic quantum criticality \cite{BKV,Pepin_FM,Chubukov_FM}, which may give rise to the first order transition. It has been explicitly demonstrated that this BKV ``instability" occurs in the case of SU(2) symmetry \cite{Pepin_FM}. In other words, if one considers Ising symmetry, it does not arise \cite{Pepin_FM}. In the U(1) slave spin-rotor formulation ferromagnetic amplitude fluctuations are described by a scalar field instead of the SO(3) vector field, where such additional components are described by spinon excitations in the CP$^{1}$ representation. In this respect the BKV instability may not arise in the U(1) slave spin-rotor theory. Of course, the nature of the ferromagnetic quantum phase transition is not completely clear in the U(1) slave spin-rotor theory, where gauge fluctuations may be responsible for the first order behavior, referred to as the Coleman-Weinberg mechanism \cite{Coleman_Weinberg} or the fluctuation-induced first-order phase transition \cite{Fluctuation_Induced_First_Order}.

\subsection{Considering amplitude fluctuations only}

We perform the scaling analysis for our renormalized field theory Eq. (\ref{U1SSR_EFT}). Before we take into account all terms of this effective field theory, we focus on longitudinal spin fluctuations first in order to set our reference which corresponds to the Hertz-Moriya-Millis theory. Consider the field theory given by \bqa && Z = \int D f_{s\sigma} D \phi D a \exp\Bigl[ - \int_{0}^{\beta} d \tau \int_{-\infty}^{\infty} d x \int_{-\infty}^{\infty} d y \Bigl\{ f_{s\sigma}^{\dagger} \Bigl(- i \frac{c}{N_{\sigma}} (- \partial_{\tau}^{2})^{\frac{1}{3}} - i s v_{F} \partial_{x} - \frac{v_{F}}{2\gamma} \partial_{y}^{2} \Bigr) f_{s\sigma} \nn && + \phi \Bigl( \gamma \frac{\sqrt{- \partial_{\tau}^{2}}}{\sqrt{- \partial_{y}^{2}}} - v_{\phi}^{2} \partial_{y}^{2} \Bigr) \phi + \frac{u_{\phi}}{2} \phi^{4} + a \Bigl( \gamma \frac{\sqrt{- \partial_{\tau}^{2}}}{\sqrt{- \partial_{y}^{2}}} - v_{a}^{2} \partial_{y}^{2} \Bigr) a - \frac{g_{\phi}}{\sqrt{N_{\sigma}}} \phi \sigma f_{s\sigma}^{\dagger} f_{s\sigma} - \frac{e_{f}}{\sqrt{N_{\sigma}}} s v_{F} a \sigma f_{s\sigma}^{\dagger} f_{s\sigma} \Bigr\} \Bigr] . \label{HMM_CFT} \eqa Performing the Fourier transformation, we obtain \bqa && Z = \int D f_{s\sigma} D \phi D a \exp\Bigl[ - \int_{-\infty}^{\infty} \frac{d \omega}{2\pi} \int_{-\infty}^{\infty} \frac{d k_{\parallel}}{2\pi} \int_{-\infty}^{\infty} \frac{d k_{\perp}}{2\pi} \Bigl\{ f_{s\sigma}^{\dagger}(\omega,k_{\parallel},k_{\perp}) \Bigl(- i \frac{c ~ \mbox{sgn}(\omega)}{N_{\sigma}} |\omega|^{\frac{2}{3}} - s v_{F} k_{\parallel} - \frac{v_{F}}{2\gamma} k_{\perp}^{2} \Bigr) f_{s\sigma}(\omega,k_{\parallel},k_{\perp}) \nn && + \phi(\omega,k_{\parallel},k_{\perp}) \Bigl( \gamma \frac{|\omega|}{|k_{\perp}|} + v_{\phi}^{2} k_{\perp}^{2} \Bigr) \phi(-\omega,-k_{\parallel},-k_{\perp}) + a(\omega,k_{\parallel},k_{\perp}) \Bigl( \gamma \frac{|\omega|}{|\bm{k}_{\perp}|} + v_{a}^{2} k_{\perp}^{2} \Bigr) a(-\omega,-k_{\parallel},-k_{\perp}) \nn && - \int_{-\infty}^{\infty} \frac{d \Omega}{2\pi} \int_{-\infty}^{\infty} \frac{d q_{\parallel}}{2\pi} \int_{-\infty}^{\infty} \frac{d q_{\perp}}{2\pi} \Bigl( \frac{g_{\phi}}{\sqrt{N_{\sigma}}} \phi(\Omega,q_{\parallel},q_{\perp}) \sigma f_{s\sigma}^{\dagger}(\omega+\Omega,k_{\parallel}+q_{\parallel},k_{\perp}+q_{\perp}) f_{s\sigma}(\omega,k_{\parallel},k_{\perp}) \nn && + \frac{e_{f}}{\sqrt{N_{\sigma}}} s a(\Omega,q_{\parallel},q_{\perp}) v_{F} \sigma f_{s\sigma}^{\dagger}(\omega+\Omega,k_{\parallel}+q_{\parallel},k_{\perp}+q_{\perp}) f_{s\sigma}(\omega,k_{\parallel},k_{\perp}) \Bigr) \Bigr\} \Bigr] . \eqa

Assuming the robustness of fermion dynamics, we introduce the scale transformation of \bqa && \omega = b^{-1} \omega' , ~~~~~ k_{\parallel} = b^{-\frac{2}{3}} k_{\parallel}' , ~~~~~ k_{\perp} = b^{-\frac{1}{3}} k_{\perp}' , \eqa which leads all renormalized kinetic energies of holons, longitudinal spin fluctuations, and U(1) gauge fluctuations to be invariant under the transformation of \bqa && f_{s\sigma}(\omega,k_{\parallel},k_{\perp}) = b^{\frac{4}{3}} f_{s\sigma}'(\omega',k_{\parallel}',k_{\perp}') , ~~~~~ \phi(\omega,k_{\parallel},k_{\perp}) = b^{\frac{4}{3}} \phi'(\omega',k_{\parallel}',k_{\perp}') , ~~~~~ a(\omega,k_{\parallel},k_{\perp}) = b^{\frac{4}{3}} a'(\omega',k_{\parallel}',k_{\perp}') . \eqa
Then, both the spin-fermion coupling and holon-gauge interaction are marginal in two dimensions, shown from \bqa && g_{\phi} = b^{-\frac{d-2}{6}} g_{\phi}' , ~~~~~ e_{f} = b^{-\frac{d-2}{6}} e_{f}' \eqa in $d-$dimensions. Eq. (\ref{HMM_CFT}) is a critical field theory, which corresponds to the Hertz-Moriya-Millis theory.

\subsection{Considering the bosonic sector only}

It is interesting to focus on the bosonic sector, given by \bqa && Z = \int D z_{\sigma} D \phi D a \exp\Bigl[ - \int_{0}^{\beta} d \tau \int_{-\infty}^{\infty} d x \int_{-\infty}^{\infty} d y \Bigl\{ \phi \Bigl( \gamma \frac{\sqrt{- \partial_{\tau}^{2}}}{\sqrt{- \partial_{x}^{2} - \partial_{y}^{2}}} - v_{\phi}^{2} \partial_{x}^{2} - v_{\phi}^{2} \partial_{y}^{2} \Bigr) \phi + \frac{u_{\phi}}{2} \phi^{4} \nn && + a \Bigl( \gamma \frac{\sqrt{- \partial_{\tau}^{2}}}{\sqrt{- \partial_{x}^{2} - \partial_{y}^{2}}} - v_{a}^{2} \partial_{x}^{2} - v_{a}^{2} \partial_{y}^{2} \Bigr) a - \frac{g_{d}}{N_{\sigma}} ( z_{\sigma}^{\dagger} \partial_{\tau} z_{\sigma} ) \frac{1}{\gamma \frac{\sqrt{- \partial_{\tau}^{2}}}{\sqrt{- \partial_{x}^{2} - \partial_{y}^{2}}} - v_{\phi}^{2} \partial_{x}^{2} - v_{\phi}^{2} \partial_{y}^{2}} ( z_{\sigma'}^{\dagger} \partial_{\tau} z_{\sigma'} ) \nn && + z_{\sigma}^{\dagger} ( - v_{z}^{2} \partial_{x}^{2} - v_{z}^{2} \partial_{y}^{2} + m_{z}^{2} ) z_{\sigma} + \frac{u_{z}}{2} |z_{\sigma}|^{4} - \frac{g_{z}}{\sqrt{N_{\sigma}}} \phi z_{\sigma}^{\dagger} \partial_{\tau} z_{\sigma} - i \frac{e_{z}}{\sqrt{N_{\sigma}}} a [ z_{\sigma}^{\dagger} (\partial_{x} z_{\sigma}) - (\partial_{x} z_{\sigma}^{\dagger}) z_{\sigma}] \Bigr\} \Bigr] , \eqa where $-\partial_{x}^{2}$ has been introduced since there does not exist a Fermi surface in this consideration. Performing the Fourier transformation, we obtain \bqa && Z = \int D z_{\sigma} D \phi D a \exp\Bigl[ - \int_{-\infty}^{\infty} \frac{d \omega}{2\pi} \int_{-\infty}^{\infty} \frac{d k_{\parallel}}{2\pi} \int_{-\infty}^{\infty} \frac{d k_{\perp}}{2\pi} \Bigl\{ \phi(\omega,k_{\parallel},k_{\perp}) \Bigl( \gamma \frac{|\omega|}{\sqrt{k_{\parallel}^{2} + k_{\perp}^{2}}} + v_{\phi}^{2} k_{\parallel}^{2} + v_{\phi}^{2} k_{\perp}^{2} \Bigr) \phi(-\omega,-k_{\parallel},-k_{\perp}) \nn && + a(\omega,k_{\parallel},k_{\perp}) \Bigl( \gamma \frac{|\omega|}{\sqrt{k_{\parallel}^{2} + k_{\perp}^{2}}} + v_{a}^{2} k_{\parallel}^{2} + v_{a}^{2} k_{\perp}^{2} \Bigr) a(-\omega,-k_{\parallel},-k_{\perp}) - \frac{g_{d}}{N_{\sigma}} \int_{-\infty}^{\infty} \frac{d \omega'}{2\pi} \int_{-\infty}^{\infty} \frac{d k_{\parallel}'}{2\pi} \int_{-\infty}^{\infty} \frac{d k_{\perp}'}{2\pi} \int_{-\infty}^{\infty} \frac{d \Omega}{2\pi} \int_{-\infty}^{\infty} \frac{d q_{\parallel}}{2\pi} \nn && \int_{-\infty}^{\infty} \frac{d q_{\perp}}{2\pi} z_{\sigma}^{\dagger}(\omega+\Omega,k_{\parallel}+q_{\parallel},k_{\perp}+q_{\perp}) z_{\sigma}(\omega,k_{\parallel},k_{\perp}) \frac{(i\omega + i\Omega/2)(i\omega' - i\Omega/2)}{\gamma \frac{|\Omega|}{\sqrt{q_{\parallel}^{2} + q_{\perp}^{2}}} + v_{\phi}^{2} q_{\parallel}^{2} + v_{\phi}^{2} q_{\perp}^{2}} z_{\sigma}^{\dagger}(\omega'-\Omega,k_{\parallel}'-q_{\parallel},k_{\perp}'-q_{\perp}) z_{\sigma}(\omega',k_{\parallel}',k_{\perp}') \nn && + z_{\sigma}^{\dagger}(\omega,k_{\parallel},k_{\perp}) (v_{z}^{2} k_{\parallel}^{2} + v_{z}^{2} k_{\perp}^{2} + m_{z}^{2}) z_{\sigma}(\omega,k_{\parallel},k_{\perp}) \nn && - \int_{-\infty}^{\infty} \frac{d \Omega}{2\pi} \int_{-\infty}^{\infty} \frac{d q_{\parallel}}{2\pi} \int_{-\infty}^{\infty} \frac{d q_{\perp}}{2\pi} \Bigl( \frac{g_{z}}{\sqrt{N_{\sigma}}} \Bigl( i \omega + \frac{i \Omega}{2} \Bigr) \phi(\Omega,q_{\parallel},q_{\perp}) z_{\sigma}^{\dagger}(\omega+\Omega,k_{\parallel}+q_{\parallel},k_{\perp}+q_{\perp}) z_{\sigma}(\omega,k_{\parallel},k_{\perp}) \nn && + \frac{e_{z}}{\sqrt{N_{\sigma}}} a(\Omega,q_{\parallel},q_{\perp}) \Bigl( k_{\parallel} + \frac{q_{\parallel}}{2} \Bigr) z_{\sigma}^{\dagger}(\omega+\Omega,k_{\parallel}+q_{\parallel},k_{\perp}+q_{\perp}) z_{\sigma}(\omega,k_{\parallel},k_{\perp}) \Bigr) \Bigr\} \Bigr] . \eqa

Hinted from the $z = 3$ critical dynamics, it is natural to take the scale transformation of \bqa && \omega = b^{-1} \omega' , ~~~~~ k_{\parallel} = b^{-\frac{1}{3}} k_{\parallel}' , ~~~~~ k_{\perp} = b^{-\frac{1}{3}} k_{\perp}' . \eqa Notice that the scale transformation for both momenta is isotropic. It is straightforward to check out that \bqa && \phi(\omega,k_{\parallel},k_{\perp}) = b^{\frac{7}{6}} \phi'(\omega',k_{\parallel}',k_{\perp}') , ~~~~~ a(\omega,k_{\parallel},k_{\perp}) = b^{\frac{7}{6}} a'(\omega',k_{\parallel}',k_{\perp}') \eqa guarantee the scale invariance for their renormalized kinetic energies.

On the other hand, there appears uncertainty for the scale transformation in dynamics of spinons. First, we consider \bqa && z_{\sigma}(\omega,k_{\parallel},k_{\perp}) = b^{\frac{7}{6}} z_{\sigma}'(\omega',k_{\parallel}',k_{\perp}') , \eqa which guarantees the scale invariance for their kinetic energy. However, it turns out that this transformation makes all interaction vertices involved with transverse spin fluctuations relevant, shown from \bqa && g_{d} = b^{\frac{5}{3}} g_{d}', ~~~~~ g_{z} = b^{\frac{5}{3}} g_{z}', ~~~~~ e_{z} = b^{\frac{1}{6}} e_{z}' . \eqa As a result, this transformation rule does not give a fixed-point theory. We interpret this situation as follows. If we assume the scale invariance of the momentum sector in dynamics of transverse spin fluctuations, the frequency term turns out to be relevant. Then, the spinon dynamics is expected to be static at low energies since only the zero-frequency sector is allowed. As a result, spinons are forced to condense at zero temperature. The condensation of spinons leads us to return back to the Hertz-Moriya-Millis description, where gauge fluctuations become gapped due to Anderson-Higgs mechanism \cite{Many_Body_Textbook}. In this case the U(1) slave spin-rotor theory recovers the Hertz-Moriya-Millis ``fixed point".


Let us consider the second scale transformation for the spinon field, given by \bqa && z_{\sigma}(\omega,k_{\parallel},k_{\perp}) = b^{\frac{19}{12}} z_{\sigma}'(\omega',k_{\parallel}',k_{\perp}') , \eqa which leads the nonlocal temporal-correlation term invariant. This scale transformation makes both the velocity and mass of spinons irrelevant, shown by \bqa && v_{z}^{2} = b^{- \frac{5}{6}} {v_{z}^{2}}' , ~~~~~ m_{z}^{2} = b^{- \frac{3}{2}} {m_{z}^{2}}' . \eqa As a result, dynamics of transverse spin fluctuations becomes locally critical at this ferromagnetic quantum critical point, identified with a novel fixed point in the strong coupling regime of the Hertz-Moriya-Millis theory. It is straightforward to see that the interaction vertex between spinons and U(1) gauge fields is irrelevant, given by $e_{z} = b^{- \frac{2}{3}} e_{z}'$, while the spin-boson coupling is marginal, i.e., $g_{z} = g_{z}'$. As a result, we find a critical field theory
\bqa && Z = \int D z_{\sigma} D \phi D a \exp\Bigl[ - \int_{0}^{\beta} d \tau \int_{-\infty}^{\infty} d x \int_{-\infty}^{\infty} d y \Bigl\{ \phi \Bigl( \gamma \frac{\sqrt{- \partial_{\tau}^{2}}}{\sqrt{- \partial_{x}^{2} - \partial_{y}^{2}}} - v_{\phi}^{2} \partial_{x}^{2} - v_{\phi}^{2} \partial_{y}^{2} \Bigr) \phi \nn && - \frac{g_{d}}{N_{\sigma}} ( z_{\sigma}^{\dagger} \partial_{\tau} z_{\sigma} ) \frac{1}{\gamma \frac{\sqrt{- \partial_{\tau}^{2}}}{\sqrt{- \partial_{x}^{2} - \partial_{y}^{2}}} - v_{\phi}^{2} \partial_{x}^{2} - v_{\phi}^{2} \partial_{y}^{2}} ( z_{\sigma'}^{\dagger} \partial_{\tau} z_{\sigma'} ) - \frac{g_{z}}{\sqrt{N_{\sigma}}} \phi z_{\sigma}^{\dagger} \partial_{\tau} z_{\sigma} \Bigr\} \Bigr] , \eqa where longitudinal spin fluctuations remain coupled with transverse spin excitations at the ferromagnetic quantum critical point. Since the $\phi-$sector may be identified with the Hertz-Moriya-Millis theory, this critical field theory implies a novel fixed point in the strong coupling regime of the Hertz-Moriya-Millis theory, where locally critical transverse spin fluctuations are expected to modify the $z = 3$ critical physics of the Hertz-Moriya-Millis theory.

\subsection{A novel critical field theory for ferromagnetic quantum criticality}

Now, we consider Eq. (\ref{U1SSR_EFT}). We take the scale transformation given by \bqa && \omega = b^{-1} \omega' , ~~~~~ k_{\parallel} = b^{-\frac{2}{3}} k_{\parallel}' , ~~~~~ k_{\perp} = b^{-\frac{1}{3}} k_{\perp}' . \eqa Notice that the longitudinal momentum scales differently from the transverse momentum. It is straightforward to see that \bqa && f_{s\sigma}(\omega,k_{\parallel},k_{\perp}) = b^{\frac{4}{3}} f_{s\sigma}'(\omega',k_{\parallel}',k_{\perp}') , ~~~~~ \phi(\omega,k_{\parallel},k_{\perp}) = b^{\frac{4}{3}} \phi'(\omega',k_{\parallel}',k_{\perp}') , ~~~~~ a(\omega,k_{\parallel},k_{\perp}) = b^{\frac{4}{3}} a'(\omega',k_{\parallel}',k_{\perp}') \eqa lead all kinetic energies of holons, longitudinal spin fluctuations, and U(1) gauge fields invariant under the scale transformation.

Following the previous discussion, we consider \bqa && z_{\sigma}(\omega,k_{\parallel},k_{\perp}) = b^{\frac{11}{6}} z_{\sigma}'(\omega',k_{\parallel}',k_{\perp}') , \eqa which makes the nonlocal temporal-correlation term of spinons invariant. This transformation rule causes both the velocity and mass of spinons irrelevant, given by \bqa && v_{z}^{2} = b^{- 1} {v_{z}^{2}}' , ~~~~~ m_{z}^{2} = b^{- \frac{5}{3}} {m_{z}^{2}}' . \eqa It is quite interesting that all interaction vertices turn out to be marginal at this fixed point except for the spinon-gauge coupling, given by $e_{z} = b^{- \frac{2}{3}} e_{z}'$.

Finally, we find a critical field theory, identifying a novel fixed point in the strong coupling regime of the Hertz-Moriya-Millis theory, \bqa && Z = \int D f_{s\sigma} D z_{\sigma} D \phi D a \exp\Bigl[ - \int_{0}^{\beta} d \tau \int_{-\infty}^{\infty} d x \int_{-\infty}^{\infty} d y \Bigl\{ f_{s\sigma}^{\dagger} \Bigl(- i \frac{c}{N_{\sigma}} (- \partial_{\tau}^{2})^{\frac{1}{3}} - i s v_{F} \partial_{x} - \frac{v_{F}}{2\gamma} \partial_{y}^{2} \Bigr) f_{s\sigma} \nn && + \phi \Bigl( \gamma \frac{\sqrt{- \partial_{\tau}^{2}}}{\sqrt{- \partial_{y}^{2}}} - v_{\phi}^{2} \partial_{y}^{2} \Bigr) \phi - \frac{g_{\phi}}{\sqrt{N_{\sigma}}} \phi \sigma f_{s\sigma}^{\dagger} f_{s\sigma} + a \Bigl( \gamma \frac{\sqrt{- \partial_{\tau}^{2}}}{\sqrt{- \partial_{y}^{2}}} - v_{a}^{2} \partial_{y}^{2} \Bigr) a - \frac{e_{f}}{\sqrt{N_{\sigma}}} s v_{F} a \sigma f_{s\sigma}^{\dagger} f_{s\sigma} \nn && - \frac{g_{d}}{N_{\sigma}} ( z_{\sigma}^{\dagger} \partial_{\tau} z_{\sigma} ) \frac{1}{\gamma \frac{\sqrt{- \partial_{\tau}^{2}}}{\sqrt{- \partial_{y}^{2}}} - v_{\phi}^{2} \partial_{y}^{2}} ( z_{\sigma'}^{\dagger} \partial_{\tau} z_{\sigma'} ) - \frac{g_{c}}{N_{\sigma}} \sigma f_{s\sigma}^{\dagger} f_{s\sigma} \frac{1}{\gamma \frac{\sqrt{- \partial_{\tau}^{2}}}{\sqrt{- \partial_{y}^{2}}} - v_{\phi}^{2} \partial_{y}^{2}} (z_{\sigma'}^{\dagger} \partial_{\tau} z_{\sigma'}) - \frac{g_{z}}{\sqrt{N_{\sigma}}} \phi z_{\sigma}^{\dagger} \partial_{\tau} z_{\sigma} \Bigr\} \Bigr] . \label{U1SSR_CFT} \eqa

\begin{figure}[t]
\includegraphics[width=0.8\textwidth]{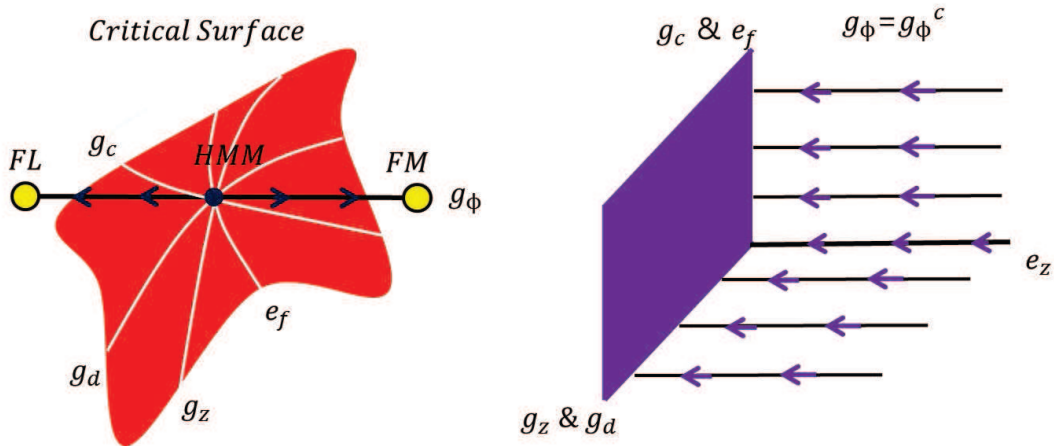}
\caption{A schematic diagram for renormalization group flows in a ferromagnetic quantum phase transition. The U(1) slave spin-rotor theory suggests possible existence of another fixed point beyond the Hertz-Moriya-Millis theory on the ``Hertz-Moriya-Millis" critical surface. When spinons are forced to condense, the Hertz-Moriya-Millis fixed point would be realized as discussed before. On the other hand, if dynamics of such transverse spin fluctuations becomes localized, we expect to reach the U(1) slave spin-rotor fixed point [Eq. (\ref{U1SSR_CFT})], differentiated from the Hertz-Moriya-Millis.} \label{RG_Flow_FMQCP}
\end{figure}

It is straightforward to extend the present analysis into the three dimensional case, where the fermion self-energy is proportional to $|\omega|$ linearly. If the $z = 3$ quantum criticality of longitudinal spin fluctuations is forced to be protected, the $- \bm{\partial}_{\bm{y}}^{2}$ term in the dispersion of holons turns out to be irrelevant, which should be backup with $|- \bm{\partial}_{\bm{y}}^{2}|^{3/2}$ as the next leading order for the curvature term. Then, the $z = 3$ quantum criticality leads all interactions marginal except for the spin-gauge coupling vertex denoted by $e_{z}$, irrelevant. As a result, we reach essentially the same expression as Eq. (\ref{U1SSR_CFT}). The physical picture of our renormalization group analysis is presented in Fig. \ref{RG_Flow_FMQCP}.

\subsection{Prediction}

It is not difficult to show that the uniform spin susceptibility contributed from order parameter fluctuations, given by a convolution integral of the amplitude-fluctuation ($\phi$) and directional-fluctuation ($z_{\sigma}$) propagators, i.e., $\langle \bm{\Phi}^{+}(\bm{r},\tau) \bm{\Phi}^{-}(\bm{r}',\tau') \rangle \sim \langle \phi(\bm{r},\tau) \phi(\bm{r}',\tau') \rangle \langle z_{\uparrow}^{\dagger}(\bm{r},\tau) z_{\downarrow}(\bm{r},\tau) z_{\downarrow}^{\dagger}(\bm{r}',\tau') z_{\uparrow}(\bm{r}',\tau') \rangle$, is proportional to $1/T$, the Curie-like behavior due to the contribution from emergent localized directional spin fluctuations (spinons). However, it is not clear whether the $1/T$ behavior is preserved beyond this level of approximation, where we expect $1/T^{1 + \eta}$ with an exponent $\eta$. In particular, we speculate that the emergent local quantum criticality, if it exists indeed, will allow the $\omega / T$ or $H / T$ scaling physics, where $H$ is a magnetic field. As a result, we propose \bqa && \chi_{u}(H,T) = \frac{1}{T^{1 + \eta}} F\Bigl(\frac{H}{T}\Bigr) , \eqa where $F(H/T)$ is a scaling function.

\section{Summary and discussion}

Emergence of localized magnetic moments and their role in metal-insulator transitions have been central issues for strongly correlated electrons. In the present study we demonstrated that such localized magnetic moments can appear in magnetic quantum phase transitions of itinerant electrons. Of course, the interpretation for the emergence of localized magnetic moments at ferromagnetic quantum criticality should be checked out more carefully, where only transverse spin fluctuations are locally critical but the correlation length in longitudinal spin fluctuations is still diverging. However, it looks plausible that the strong coupling regime may not be described by the Hertz-Moriya-Millis theory. Instead, dynamics of spin fluctuations can be modified, here localized for transverse spin fluctuations due to strong correlations with both longitudinal spin fluctuations and itinerant electrons, while the dynamics of longitudinal spin fluctuations is still of Hertz-Moriya-Millis at least in the Eliashberg approximation. We believe that the emergence of localized magnetic moments at quantum criticality is not limited in ferromagnetism. Recently, we investigated an antiferromagnetic quantum phase transition with an ordering wave vector $2 \bm{k}_{F}$ based on the U(1) slave spin-rotor representation of the Hertz-Moriya-Millis theory, regarded to be essentially the same strong coupling approach as the present study \cite{Kim_U1SSR_AFQCP}. There, we found that not only transverse spin fluctuations but also longitudinal spin fluctuations are locally critical, implying that this novel fixed point is described by a critical field theory in terms of emergent locally critical magnetic moments and renormalized electrons. Since dynamics of transverse spin excitations is locally critical, i.e., impurity-like, we expect the $\omega/T$ scaling physics beyond the weak coupling regime of the Hertz-Moriya-Millis theory. This $\omega/T$ scaling physics should be investigated more sincerely near future.

\section*{Acknowledgement}

This study was supported by the Ministry of Education, Science, and Technology (No. 2012R1A1B3000550) of the National Research Foundation of Korea (NRF) and by TJ Park Science Fellowship of the POSCO TJ Park Foundation.

\end{document}